\documentclass[aps,pra,showkeys,twocolumn]{revtex4}

\usepackage{graphicx}
\begin{document}

\title{Quasi-determinism of weak measurement statistics: \\
Laplace's demon's quantum cousin}

\author{Holger F. Hofmann}
\email{hofmann@hiroshima-u.ac.jp}
\affiliation{
Graduate School of Advanced Sciences of Matter, Hiroshima University,
Kagamiyama 1-3-1, Higashi Hiroshima 739-8530, Japan}
\affiliation{JST,CREST, Sanbancho 5, Chiyoda-ku, Tokyo 102-0075, Japan
}

\begin{abstract}
Weak measurements can provide a complete characterization of post-selected ensembles, including the uncertainties of observables. Interestingly, the average uncertainties for pure initial and final states are always zero, suggesting the kind of complete knowledge that would allow a knowledge of past, presence and future in the sense of Laplace's demon. However, the quantum version actually describes cancellations of positive and negative uncertainties made possible by the strangeness of weak values. In this paper, I take a closer look at the relation between statistics and causality in quantum mechanics, in an attempt to recover the traces of classical determinism in the statistical relations of quantum measurement outcomes.
\end{abstract}


\maketitle

\section{Introduction}
In 1814, Pierre-Simon Laplace argued in ``A Philosophical Essay on Probabilities'' that an entity knowing the present state of the universe would also know its past and future, without any uncertainty. Laplace's speculation is clearly rooted in the well-defined description of causal determinism provided by Newtonian mechanics, where it corresponds to the equations of motion that deterministically transform initial states into final states. In quantum mechanics, the uncertainty principle seems to remove the foundations on which the argument rests. However, the uncertainty principle does not apply to the causality between initial and final states, only to the incomplete information provided by these states. The paradox of quantum mechanics is that it completely preserves the deterministic structure of classical physics while denying the possibility of complete knowledge. The reasons for this impossibility of complete knowledge cannot be explained by the unitary evolution of quantum states, since unitary evolutions preserve the available information and are therefore as deterministic as Newtonian trajectories. Instead, it is necessary to take a closer look at the relation between statistical statements and empirical facts in quantum mechanics.   

\section{Causality relations for weak values}
Empirical knowledge is not limited to predictions made about the yet unknown future. A complete description of a physical system should include all of the information available after all measurement processes are completed. In quantum mechanics, this means that equal weight should be given to initial information about the state prepared and final information about the results of all measurements performed on the system. The conventional predictive quantum formalism overemphasizes state preparation, and therefore represents only an incomplete description of empirical reality. In principle, any non-trivial final measurement adds new information, so that the complete description of the process exceeds the uncertainty limits that apply to possible predictions. In fact, classical physics indicates that initial preparation and final measurement can provide enough information for a complete reconstruction of all physical properties of the system. The uncertainty principle therefore does not limit the amount of information obtained about an individual system. It merely states that there cannot be an experimental test of any reconstructed properties, because any additional measurement performed between initial preparation and final measurement will disturb the dynamics. However, the uncertainty principle is not an all-or-nothing rule since there is a quantitative trade-off between the disturbance and the measurement resolution. In 1988, Aharonov, Albert and Vaidman discovered that the statistical averages of the outcomes can be evaluated even when the measurement interaction (and hence the disturbance) tends to zero [1]. The averaged outcomes of such weak measurements of an observable $\hat{A}$ for a fixed initial state $\mid i \rangle$ and a fixed final state $\mid f \rangle$ are described by the real part of the weak values,
\begin{equation}
\langle \hat{A} \rangle_{\mathrm{weak}} = \frac{\langle f \mid \hat{A} \mid i \rangle}
{\langle f \mid i \rangle}.
\label{eq:weakvalue}
\end{equation} 
Interestingly, weak values can actually confirm some basic expectations of classical causality. 
If either $\mid i \rangle$ or $\mid f \rangle$ is an eigenstate of $\hat{A}$, the weak value is just the corresponding eigenvalue of the initial or final state. Since the weak value of $\hat{A}=\hat{X}+\hat{Y}$ is equal to the sum of the weak values of $\hat{X}$ and $\hat{Y}$, this means that the weak value of $\hat{X}+\hat{Y}$ for an initial eigenstate of $\hat{X}$ and a final eigenstate of $\hat{Y}$ is equal to the sum of the eigenvalues of $\hat{X}$ and $\hat{Y}$. Ironically, this consistency with classical expectations gives rise to the fundamental paradox of weak values, namely that they can lie outside the range of eigenvalues of the operators representing the observables [1]. Just how paradoxical the combination of quantization and causality is can be illustrated by a simple example. In the case where $\hat{X}$ and $\hat{Y}$ are Pauli operators, the sum of the eigenvalues $X=1$ and $Y=1$ is $X+Y=2$. However, this is clearly larger than the maximal eigenvalue of $\hat{X}+\hat{Y}$, which is only $\sqrt{2}$. Thus, weak measurements indicate that causality may be valid despite the fact that the eigenvalues of $\hat{X}+\hat{Y}$ are not equal to the sums of the eigenvalues of $\hat{X}$ and $\hat{Y}$. It may therefore be a mistake to hastily dismiss causality when faced with the seemingly fundamental randomness of quantum statistics. Instead, one should keep in mind that the quantized eigenvalues observed in different measurements fail to provide a consistent description of measurement independent reality, and that all measurement process are still based on the (classical) assumption of causality connecting the registered signal to the target observable.

\section{Complete quantum statistics of post-selected weak measurements}

Experimentally, weak values can only be obtained as averages over a large number of post-selected measurements. In this sense, weak values are clearly non-deterministic: for a single event, the assignment of a weak value has no clear physical meaning. On the other hand, the post-selection of an outcome $\mid f \rangle$ selects a well-defined sub-ensemble of the initial state $\mid i \rangle$. The quantum statistical properties of this sub-ensemble can be summarized by the transient density operator $\hat{R}_{if}$, so that the initial state is a mixture of post-selected ensembles given by
\begin{equation}
\mid i \rangle \langle i \mid = \sum_f
|\langle f \mid i \rangle|^2 \hat{R}_{if}.
\label{eq:decomp}
\end{equation}
As shown in [2], $\hat{R}_{if}$ can be determined from a sufficient set of measurement results using standard quantum tomography. In this case, $\hat{R}_{if}$ is a self-adjoint operator with real eigenvalues, corresponding to the fact that the self-adjoint measurement operators represent only the real parts of the weak values given by eq.(\ref{eq:weakvalue}). To include the imaginary parts of the weak values, $\hat{R}_{if}$ can be defined as the ordered product of the projectors, normalized to have a trace of one, 
\begin{equation}
\hat{R}_{if} = \frac{1}{|\langle f \mid i \rangle|^2}\left(\mid i \, \rangle \langle \, i \mid f \rangle \langle f \mid \right).
\label{eq:suben}
\end{equation}
Eqs. (\ref{eq:decomp}) and (\ref{eq:suben}) explain how the final measurement determines the quantum statistics of the system before the measurement by selecting a statistical sub-ensemble of the initial state. In close analogy to classical statistics, it is then possible to formulate an expression for the conditional probabilities for the eigenvalues of an observable $\hat{A}$. These conditional probabilities correspond to the weak values of the projection on the corresponding eigenstates $\mid A \rangle$,
\begin{equation}
p_{\mathrm{w.}}(A|f) = \frac{\langle A \mid i \, \rangle \langle \, i \mid f \rangle 
\langle f \mid A \rangle}{|\langle f \mid i \rangle|^2}
\label{eq:prob}
\end{equation}
As recently pointed out by Hosoya [3], the weak conditional probability given by eq. (\ref{eq:prob}) may provide an important link between the quantized eigenvalues of $\hat{A}$ observed in strong measurements and the distribution of weak values
observed for the possible post-selected outcomes $f$. In particular, it is now possible to consider the relation between the uncertainty of $\hat{A}$ in the initial state $\mid i \rangle$ and the conditional uncertainties obtained for each final state $\mid f \rangle$.

\section{Weak value uncertainties}

The weak values $\langle A \rangle_{\mathrm{w.}}(f)$ can now be interpreted as the average values of $\hat{A}$ in the sub-ensemble $\hat{R}_{if}$ described by the probability distribution $p_{\mathrm{w.}}(A|f)$. In general, the values of $\hat{A}$ fluctuate in each sub-ensemble, as given by the conditional uncertainty,
\begin{equation}
\Delta A^2_{\mathrm{w.}}(f) = \langle \hat{A}^2 \rangle_{\mathrm{w.}} (f) - 
|\langle \hat{A} \rangle_{\mathrm{w.}}(f)|^2. 
\end{equation} 
The initial uncertainty $\Delta A^2$ for the total ensemble $\mid i \rangle \langle i \mid$ is equal to the weighted average of the conditional uncertainties plus the absolute square of the differences between the weak value and the average value of the total ensemble, 
\begin{eqnarray}
\lefteqn{
\Delta A^2 =} \nonumber \\ && \sum_f p(f) \left(\Delta A_{\mathrm{w.}}^2 (f) + 
\left(\langle \hat{A} \rangle_{\mathrm{w.}}(f) - \langle \hat{A} \rangle \right)^2\right).
\nonumber \\
\end{eqnarray}
This means that the average conditional uncertainty $\Delta A_{\mathrm{w.}}^2 (f)$ is given by the difference between the initial uncertainty $\Delta A^2$ and the variance of the weak values.
Using eq.(\ref{eq:weakvalue}) and the completeness relation $\sum \mid f \rangle\langle f \mid=1$,
the variance of the weak values is
\begin{eqnarray}
\lefteqn{\sum_f p(f) 
\left(\langle \hat{A} \rangle_{\mathrm{w.}}(f) - \langle \hat{A} \rangle \right)^2}
\nonumber \\ &=& (\sum_f 
\langle i \mid \hat{A} \mid f \rangle \langle f \mid \hat{A} \mid i \rangle ) - \langle \hat{A} \rangle^2
\nonumber \\ &=& \langle \hat{A}^2 \rangle  - \langle \hat{A} \rangle^2 \; = \; \; \Delta A^2.
\end{eqnarray}
Therefore, the fluctuations of $\hat{A}$ in $\mid i \rangle$ observed in precise measurements of $\hat{A}$ correspond to the fluctuations of the weak values observed for different measurement outcomes $f$. This means that the average conditional uncertainty of the transient states $\hat{R}_{if}$ must be zero,
\begin{equation}
\sum_f p(f) \Delta A_{\mathrm{w.}}^2 (f) = 0.
\label{eq:uncertaintymean}
\end{equation}
This result can also be confirmed by directly evaluating the statistics of $\hat{R}_{if}$. However, the derivation from the fluctuations
 of the weak values observed for different final results $f$ provides a more complete and experimentally accessible picture of the rather surprising observation that the average uncertainty can be zero, even though the spectrum of weak values is different from the spectrum of the eigenvalues. 

\section{Quasi-determinism}

In classical statistics, an average uncertainty of zero means that the value of the observable is known with precision. However, weak values are the averages of the non-positive probability distributions $p_{\mathrm{w.}}(A|f)$. This means that negative uncertainties are possible. In fact, the difference between the weak values and the eigenvalues of the observable $\hat{A}$ indicate that the weak value cannot be identified with a precise value of $\hat{A}$. Eq. (\ref{eq:uncertaintymean}) therefore merely indicates that the positive and negative conditional uncertainties cancel out on average. Ultimately, the relation between the weak value of $\hat{A}=\hat{X}+\hat{Y}$ and the precise values of $\hat{X}$ and $\hat{Y}$ defined by an initial eigenstate of $\hat{X}$ and a final eigenstate of $\hat{Y}$ is only quasi-deterministic, since weak values can never be more than statistical averages. On the other hand, quasi-determinism completely removes the uncertainty limit, opening up the possibility of analyzing the quantum statistical relations between non-commuting properties in more detail.

In particular, quasi-determinism offers an intriguing new perspective on the problem presented by the observation that the eigenvalues of $\hat{X}+\hat{Y}$ are fundamentally different from the possible sums of eigenvalues of $\hat{X}$ and eigenvalues of $\hat{Y}$. Simultaneous knowledge of $X$ and $Y$ does determine the average of $X+Y$, but there must still be a positive or negative uncertainty associated with the fact that $X+Y$ is not part of the eigenvalue spectrum of $\hat{X}+\hat{Y}$. For Pauli operators, the values of $X+Y$ are $0$ or $\pm 2$, but the eigenvalues of $\hat{X}+\hat{Y}$ are $\pm \sqrt{2}$. Therefore, the uncertainty of $\hat{X}+\hat{Y}$ is either $2$ (for $X+Y=0$) or $-2$ (for $X+Y=\pm 2$). Thus the average uncertainty would be zero, confirming the completeness of information provided by simultaneous knowledge of $\hat{X}$ and $\hat{Y}$, while maintaining the consistency between the quantized eigenvalues of $\hat{X}+\hat{Y}$ and of $\hat{X}$ and $\hat{Y}$ through the quantum statistical connection between the observed properties and the unobserved properties represented by non-positive probabilities. 

It might even be possible to argue that the quasi-determinism of non-positive statistics is a necessary consequence of the contradiction between the observation of discrete eigenvalues in strong measurements and the continuous causality of unitary transformations describing the evolution of physical properties. As Schr\"odinger pointed out in the first part of his famous paper on the present situation in quantum mechanics of 1935 [4], it is the inconsistency of the eigenvalues with the roles of the observables ascribed to them by the (classical) model of the system that frustrates any attempt to assign classical realities to the physical properties of a quantum system. Interestingly, quasi-determinism offers a way out that explains why the classical model still plays the role it does in the definition of the observables. In some sense, quasi-determinism is causality without realism. It can therefore preserve the structure of our classical model (and much of the classical intuition that comes with it), while accommodating the strangeness of the quantized measurement outcomes that seem to contradict the very assumptions used in the formulation of the theory.

\section{Conclusions}

In conclusion, quantum mechanics does have a well defined statistical structure beyond the uncertainty limit. In fact, complete knowledge of the history of an object implies complete knowledge of the unobserved aspects of this history, based on the causality relations between the actual observations and all of the unobserved properties of a system. However, quantization requires that the connection is not completely deterministic. Instead, quantum coherences provide non-positive joint probabilities that can attribute a negative uncertainty to an average value that does not correspond to an eigenvalue. Interestingly, the average uncertainty obtained for ensembles defined by both their initial and final conditions is still zero, demonstrating that the laws of causality enforce a quantum statistical version of determinism that I have labeled quasi-determinism. 

Quasi-determinism translates the statistical rules implied by the relations between non-orthogonal states in Hilbert space into statistical rules that correspond to the causality relations of classical physics. Ultimately, these causality relations may provide a more direct description of quantum statistics and quantum information by reconciling the discrete statistics of measurement outcomes with the continuous transformation of quantum states. The fundamental relations revealed by such an analysis could provide a unifying perspective on the wide variety of quantum statistical oddities investigated in quantum information and quantum foundations. Thus, Laplace's demon might still have something to teach us, even in the highly uncertain terrain of quantum physics.

\section*{Acknowledgements}
Part of this work has been supported by the Grant-in-Aid program of the
Japanese Society for the Promotion of Science, JSPS


\vspace{0.8cm}

\section*{References}
\noindent
[1] Y. Aharonov, D. Z. Albert, and L. Vaidman, Phys. Rev. Lett. {\bf 60}, 1351 (1988).
\\[0.1cm]
[2] H.F. Hofmann, Phys. Rev. A {\bf 81}, 012103 (2010).
\\[0.1cm]
[3] A. Hosoya, private communication and presentation at the Nagoya Winter Workshop, held February 18th to 23rd 2010 at Nagoya University, Nagoya, Japan.
\\[0.1cm]
[4] E. Schr\"odinger, Naturwissenschaften {\bf 23}, 807 (1935).
\end{document}